# Formalizing Regression Testing for Agile and Continuous Integration Environments

Suddhasvatta Das[1] and Kevin Gary[2]

[1,2]School Of Computing and Augmented Intelligence, Arizona State University, 699 South Mill Avenue, Tempe, 85281, Arizona, USA.

Contributing authors: sdas76@asu.edu; kgary@asu.edu;

**Abstract**

Software developed using modern agile practices delivers a stream of software versions that require continuous regression testing rather than testing once close to the delivery or maintenance phase, as assumed by classical regression-testing theory. In this work, we formalize the phenomenon of continuous or near-continuous regression testing using successive 'builds' as a time-ordered chain, where each build contains the program, requirements, and the accompanying tests. We also formalize the regression test window between any two builds, which captures the limited time budget available for regression testing. As the time limit is set to infinity and the chain is closed to two builds, the model degenerates to retest-all, thereby preserving semantics for the classical two-version case. The formalization is validated by directly representing two state-of-the-art agile regression testing algorithms in terms of build-tuple operations without requiring auxiliary assumptions, followed by proof of the soundness and completeness of our formalization.

## 1 Introduction

This work aims to formalize regression testing in agile environments. Agile methods continue to grow in effectiveness and popularity, with over 71% of organizations using agile as their primary software development process 3% of teams use Scrum, and 26% report using the Scaled Agile Framework (SAFe) at the enterprise level.[1]. However, with all its benefits come significant challenges, particularly around testing, and for this work, we focus specifically on regression testing. Organizations like Microsoft and Google trigger hundreds of builds daily, executing thousands of regression tests automatically before code is shipped [2]. However, the summer of 2024 *CrowdStrike* incident underscored how relying solely on automation for speedy delivery can be disastrous [3][4]. Although automation is necessary for executing regression testing, numerous modern tools and technologies are available to facilitate it, such as *Selenium, TestComplete*, and many others. Automating and/or developing AI/ML-driven algorithms [5] is not always the solution to regression testing problems in agile. Given that the fundamental objective of regression testing remains consistent across development processes, i.e., modifications do not break existing functionality [6]. Agile methods introduce a different problem; i.e., *regression testing must be executed continuously or nearly continuously rather than as a one-time activity near delivery* [5].

This concept of continuous, ongoing regression testing has been acknowledged by recent studies on regression testing in agile settings, such as Spieker et al. 2017 [7] and Bertolino et al. 2020 [8], and many others as reported by Pan et al. 2022 and Das and Gary 2025 [5] [9]. However, despite this impressive array of approaches, the field still *lacks a precise definition of the regression testing problem* rather than the techniques used to address the problem (such as selection and prioritization) in agile development, which can be characterized as a stream of continuous or near-continuous and ever-changing streams of changes versus isolated activities close to the end of development (as a maintenance activity), as shown by Do et al. 2006 [10].

In the absence of a formal definition, the regression testing definition in agile is purely a narrative. A rigorous mathematical formulation would identify and quantify the fundamental components associated with regression testing in agile, thereby transforming the phenomenon into a directly measurable entity, not just the quality metrics, such as coverage and defect detection rates of tests used in regression testing. From this definition, one can derive fundamental limits, such as *minimal resource requirements and inherent trade-offs between targeted quality metrics and effort,* without reference to any particular tool or heuristic. The formal definition can be used to derive trade-offs between *time/resource/cost vs quality* (as per set by the team/organization e.g. coverage/defect detection rate) in dynamic development cycles for a given change. In practice, this allows project managers to set concrete resource capacity benchmarks (e.g., environment setup and result inspection overhead requires specific compute hours and staffing), run scenario analyses to anticipate how changes in test suite size or test execution frequency affect feedback-loop duration.

Equipped with a precise definition, teams can also perform *"what-if"* analyses to predict how shifts in the scope of change, release cadence, or testing resources and time impact a delivery plan. Finally, by providing a shared formalism, this definition would unite researchers and practitioners under a common language, ensuring that benchmarks, tools, and theoretical advances all address precisely the same phenomenon. This work addresses the gap by extending the two-build formalization of regression testing by Rothermel and Harrold 1997 [6] to a continuous sequence of builds for regression testing for agile environments, thereby capturing the frequency, various time constraints, and dynamic evolution of requirements and tests inherent to agile methods.

Agile methodologies encompass a range of practices, broadly classified into time-boxed and feature-boxed frameworks, with Scrum and Kanban being prominent examples. Scrum follows a time-boxed approach, where iterations, known as *sprints*, are fixed in duration and typically last 2–4 weeks [5]. In contrast, Kanban employs a feature-boxed model, where work progresses continuously without fixed time constraints, prioritizing delivery once a feature is completed [11]. Software is continually delivered in both cases as code and test sets/suites evolve. While regression tests are executed in both cases, there are other activities such as environment *setup time* and checking the *results of test execution* [10] that require time and thus impact regression testing's objective and trade-offs. While automation is prevalent in established agile organizations, as stated previously, it may address many of these challenges, but automating and maintaining automation is expensive for complex enterprise business scenarios.

Thus, combining the above-mentioned aspects related to the time spent on various activities that directly or indirectly affect regression testing in agile environments adds complexity to the regression testing problem space in agile environments. Addressing this requires extending the classical regression testing definition by Rothermel and Harrold 1997 [6] for an agile environment to capture regression testing in a continuous or near-continuous environment of changes requiring regression testing with respect to time. By formally extending the foundational definitions to encompass these aspects, our work provides a systematic basis for understanding the problem space of regression testing in agile development. With this motivation, this work aims to: **Formalize Regression Testing in Agile Environments.**

The following sections provide the background for scope regression testing within the overall testing context, extend the definitions, and define agile regression testing, followed by validation of our definition of agile regression testing.

## 2 Background

This section outlines the position of regression testing within the broader testing framework and describes its current formalism in the literature. Testing is categorized as code-based or specification-based [12]. In the former, Tests are derived directly from the source code, and in the latter, specifications are used to derive tests. Irrespective of the testing, the relation between the code base $P$, specifications $S$, and test cases $T$ is illustrated in Figure 1 [12].

Regression testing focuses on region '1', ensuring that already-delivered specifications do not break with new changes. Other scenarios to consider include Region 2—specifications implemented but not tested. Region 3—implemented code with tests but no specifications, Region 4—specifications with tests but not implemented, Region 5—specifications with no implementation or testing, Region 6—implemented code without tests and specifications, and Region 7—tests without specifications and code. Region 8 encompasses all requirements, tests, and programs not associated with elements in the above *P, S*, and *T* sets.

Considering two consecutive regions, 'one' from Figure 1, that maps to program versions $P_i$ and $P_{i+1}$, let us call it "current" and "modified" as Rothermel and Harold

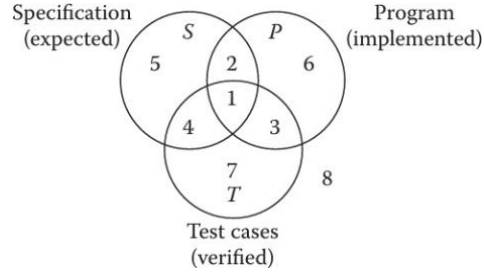

**Fig. 1** Relationship of P,S,T [12]

1997 [6]. In these two consecutive regions [see Figure 2], one represents specifications implemented from the set $S_i$ and $S_{i+1}$ and have been tested by the test sets $T_i$ and $T_{i+1}$ respectively. Between these two versions, the focus is only on the specifications that remain the same across the two versions, thus signifying that the output should be the same, the focus of regression testing with the assumption that $S_i \cap S_{i+1} \not\models \emptyset$. Therefore, the candidate set of regression testing is tests tied to these specifications, i.e., $T_i \cap T_{i+1}$. The test set $T_i \cap T_{i+1}$ represents tests carried forward from $P_i$ to validate shared specifications $S_i \cap S_{i+1}$. These tests ensure that the unchanged functionality behaves consistently across both versions. In contrast, $T_{i+1}$ includes additional tests created specifically to verify the new or modified functionality in $P_{i+1}$, corresponding to specifications $S_{i+1} - S_i$. All the other specifications are either deprecated or new, corresponding to sets $S_i - S_{i+1}$ and $S_{i+1} - S_i$, respectively, and are not the focus of regression testing.

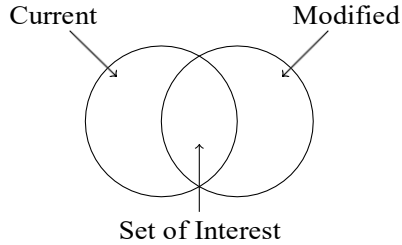

**Fig. 2** Regression Candidate

Notable works such as Rothermel and Harrold 1997 [6] and Yoo and Harman 2012 [13] have significantly contributed to formalizing regression testing. According to Rothermel and Harold 1997 [6], let $P_i$ be a program, let $P_{i+1}$ be a modified version of $P_i$, and let $S_i$ and $S_{i+1}$ be the specifications for $P_i$ and $P_{i+1}$, respectively. $P_i(\text{inp})$ refers to the output of $P_i$ on input inp, $P_{i+1}(\text{inp})$ refers to the output of $P_{i+1}$ on input inp, $S_i(\text{inp})$ refers to the specified output for $P_i$ on input inp, and $S_{i+1}(\text{inp})$ refers to the specified output for $P_{i+1}$ on input inp. Let $T_i \cap T_{i+1}$ be a set of tests created to test $P_{i+1}$. Rothermel and Harold 1997 [6] also define a test case as $t = (ID, inp, S_i(inp))$, where ID is the identifier, *inp* is the input, $S_i(inp)$ is the expected output. Thus, for simplicity we can also write $P_i$(inp) as $P_i$(t) as *inp* is a part of *t* similar to Yoo and Harman 2012 [13]. These works on formalizing regression testing provide a solid foundation for a mathematical framework that can be extended to capture agile's continuous, incremental, and iterative nature.

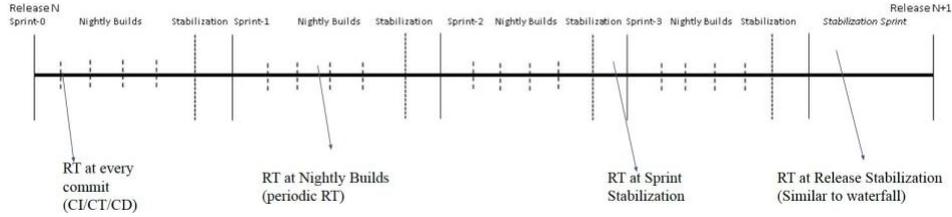

**Fig. 3** Opportunities for Regression Test in Agile

In agile, a change/update can be 'anytime' and requires regression testing, but is not limited to *feature completion time, continuous/nightly test time, sprint stabilization time, and release stabilization time* [see Fig. 3, agile cadence]. The changed/updated program can be created anytime, but is not limited to the time instances mentioned in Fig. 3. The changed program may not necessarily reach customer, but is eligible for regression testing with time limit on the amount of time available for regression testing and we call it *Regression Testing Windows* (RTW). The RTW is important in selecting and/or prioritizing test cases for regression testing [7] when all possible regression tests cannot be run.

In summary, agile development's rapid and iterative cycles introduce multiple opportunities and challenges for regression testing. Works such as those by Rothermel and Harrold 1997 [6] formalize the relationships between program versions, specifications, and test cases and provide a robust basis for structured regression testing. Similarly, Spieker et al. 2017 [7], a seminal work in the field, contributes to formalizing regression test selection and prioritization in a time-constrained agile setting. While these contributions are essential, the literature has not formally defined *regression testing as an ongoing phenomenon in a continuous agile workflow with/without time constraints*. In the next section We extend these foundational definitions to formalize this phenomenon.

## 3 Extended and New Definitions

This section extends the definitions of three components of regression testing as defined by Rothermel and Harrold's 1997 [6], i.e., programs, specifications, and test cases. We also define new items to capture agile's continuous nature namely: *builds, releases, and iterations*.

**Program Versions:** Let $\{P_1, P_2, \ldots, P_n\}$ be an ordered set of program versions, with $P_1 < P_2 < \cdots < P_n$, where $P_i < P_{i+1}$ indicates that $P_i$ occurs earlier in time than $P_{i+1}$. Each $P_i$ corresponds to a software version in the Agile development cycle, capturing all engineering done up to that point.

**Specifications:** Let $\{S_1, S_2, \ldots, S_n\}$ be an ordered set of specifications, with $S_1 < S_2 < \cdots < S_n$. Each $S_i$ corresponds to the program version $P_i$ and is represented as $S_i = \{s_j / 1 \leq j \leq q\}$, where each $s_j$ is commonly represented as a 'User Story'. In Agile, User Stories are concise descriptions of functionality required by the end-user and include attributes such as *effort and value*. Effort, often measured using 'Story Points', assesses the complexity and risk of completing a task, while value refers to the strategic or financial impact of the user story [14]. Thus, each set of specification(s) $S_i$ is a set of tuples defined as $S_i = \{(s_1, bv_1, sp_1), (s_2, bv_2, sp_2), \ldots, (s_q, bv_q, sp_q)\}$, where $(s_j, bv_j, sp_j)$ represents a users story, its business value, and its story point, with $j$ ranging from 1 to $q$, the total number of user stories in $S_i$.

**Test Sets**: Let $\{T_1, T_2, \ldots, T_n\}$ be an ordered set of test sets, with $T_1 < T_2 < \cdots < T_n$. Each $T_i$ corresponds to the program version $P_i$. Each $T_i = \{t_m / 1 \leq m \leq q\}$, where each $t_m$ is an individual test case.

We further extend the definition of a test case by explicitly incorporating the execution and setup time of each test case. Time is a critical factor in agile development settings, and previous research has already recognized test case setup and execution time as influential variables in cost-benefit analyses, such as the model proposed by Do et al. [10]. Formalizing execution and setup time as intrinsic properties of test cases is therefore essential. However, Do et al.'s model also includes additional factors external to the test case itself and thus remains outside the scope of our refined test case definition. Thus, each test case $t_m = (ID, inp, S_i(inp), t_m.exectime, t_m.setup)$, where the two new terms $t_m.exectime$ denotes the execution time, and $t_m.setup$ denotes the setup time required for test case $t_m$.

**Builds**: Let $\{B_1, B_2, \ldots, B_n\}$ be an ordered set of tuples, with $B_1 < B_2 < \cdots < B_n$. Each $B_i$ is the tuple $(P_i, S_i, T_i)$, *representing a specific increment of the soft-ware product at a given point in time*. Builds form a core component of modern agile pipelines, enabling rapid iteration and continuous integration [15]. Each build encapsulates the program code ($P$), specifications ($S$), and test cases ($T$) and is a subset of releases.

**Releases**: Let $R = \{r_1 < r_2 < \cdots < r_k\}$ are specifically designated builds that are intended for delivery to the customer (outcome of work performed during one or more iterations).

**Iterations**: Let $ITR = \{itr_1 < itr_2 < \cdots < itr_Y\}$ be time-boxed intervals within which development work is planned and executed where $Y$ is the total num- ber of iterations. A release $r_k$ is produced by a release-generation function $f_r$ as a result of a sequence of iterations, defined as $f_r : P(B) \longrightarrow R$ and each $r_k = f_r(B_{itr_j,\ldots,itr_{j+n}})$. $B_{itr_j, itr_{j+1},\ldots,itr_{j+n}}$ represents the set of builds produced during the contiguous sequence of iterations. Here, $\{itr_j, \ldots, itr_{j+n}\}$ is contiguous ordering of $itr_j < \cdots < itr_{j+n}$ of $ITR$.

### 3.1 Transition of Builds

Builds are incremented over time, transitioning from $B_i$ to $B_{i+1}$ after each iteration, where $B_i = (P_i, S_i, T_i)$ represents the tuple of program code, specifications, and test cases. These transitions are driven by changes in $P$, $S$, or $T$, and regression testing is required at each transition to ensure that unchanged functionality behaves consistently. These transitions are motivated by one or more of the following events:

*Case 1 Passage of time:* While builds can transition due to changing any of the three components in the tuple, organizations sometimes build the software for sanity. This is the most common type of build, commonly known as a periodic build; it may be created nightly or at the end of business hours.

*Case 2 New Feature:* corresponding to new specifications $S_{i+1}$ with new test cases $T_{i+1}$ and transitioning $P_i$ to $P_{i+1}$ also known as adaptive change [16]

*Case 3 Defect Fixes:* results in the change of build. In this case there is no change in $S$ and $T$ but there are changes in $P$ also known as corrective change [16].

*Case 4 Technical debt pay-down:* results in changes the $P$ and possibly $T$ but no change in $S$ also known as perfective change [16].

*Case 5 New feature but no new test:* corresponds to the case where the new feature has been implemented, but specific tests for it do not exist, but there is a change in $P$ and $S$, so there is an opportunity for regression testing.

## 3.2 Varying Regression Test Window

As builds change over time due to multiple reasons , as said in the section above, these changes can be planned or ad hoc in various instances (Figure 3), allowing for regression testing. Depending on the scope of the work, the team decides on an RTW (we are not concerned with how this RTW is set by each team, i.e., out of the scope of this work). We define this mathematically as $\Delta\tau = \tau_{i+1} - \tau_i$, where $\tau_i$ marks the point at which build $B_i$ is ready for regression testing, and $\tau_{i+1}$ marks the point at which regression testing has to be completed.

In our formalism, the RTW $\Delta\tau = \tau_{i+1} - \tau_i$ serves as a quantitative parameter that directly influences the scope of test execution. Mathematically, as $\Delta\tau$ increases, the size and scope of regression testing increase. For example, when $\Delta\tau$ is very small, corresponding to *commit* level or *pull-request* level testing, the scope of regression testing is limited. A commit is a single atomic change-set recorded in the version-control repository, while a pull request groups one or more commits for review before integration into the main branch; most CI systems launch a build for every commit or pull-request. As $\Delta\tau$ grows, more tests can be executed in nightly/periodic regression testing, and still larger values are typical of sprint and release stabilization phases.

Formally, assigning each test $t_m \in T_i \cap T_{i+1}$ a *positive cost* $c(t_m)$ equal to its execution time plus the setup time (see section 2). For a window $\Delta\tau$, define $f_{\Delta\tau}(\Delta\tau) = \max \{|T_{\Delta\tau}|: T_{\Delta\tau} \subseteq T_i \cap T_{i+1}, \sum_{t_m \in T_{\Delta\tau}} c(t_m) \leq \Delta\tau\}$, the largest number of tests whose total cost fits inside the window. Because enlarging $\Delta\tau$ cannot remove any previously feasible subset. Thus, $f_{\Delta\tau}$ is monotone i.e. if $\Delta\tau_1 < \Delta\tau_2$ then $f_{\Delta\tau}(\Delta\tau_1) \leq f_{\Delta\tau}(\Delta\tau_2)$. When $\Delta\tau$ is at least the sum of all costs in the intersection, the whole set runs and $f_{\Delta\tau}(\Delta\tau) = |T_i \cap T_{i+1}|$. This captures how time constraints govern testing scope without assuming equal test durations.

# 4 Agile Regression Testing

Given the problem setup and extended definitions up to this point, we have set the stage to represent agile regression testing formally with our extended definitions. Recalling the definition of regression testing, we aim to verify that the outcome of the specifications will stay the same after a change with the corresponding test cases of those specifications [6]. Theoretically, the entire test set $T_i \cap T_{i+1}$ should be executed to ensure confidence unchanged specifications across consecutive builds continue to behave the same, a concept known as *retest-all*. This scenario is similar to when there is no time limit and $\Delta\tau \to \infty$ [2]. We call this '*RegAll*. RegAll takes consecutive builds $B_i$ and $B_{i+1}$ and compare the outcomes of each test case in $T_i \cap T_{i+1}$ across both builds. The RegAll($B_i$, $B_{i+1}$) yields 1 if all overlapping test outcomes remain consistent between $B_i$ and $B_{i+1}$, and 0 otherwise. It assumes $f_{\Delta\tau}(\Delta\tau) = |T_i \cap T_{i+1}|$, which only holds when time is not a constraint.

Moreover, this binary result does not directly signify whether the build passes or fails. Teams extend this framework in real-world scenarios by considering the severity and complexity of defects that test cases find and aligning build decisions with pre-defined quality criteria. A separate build acceptance or release decision may still be made even if RegAll returns 0, depending on the team's acceptance criteria. Thus, mathematically we say:

$$\text{RegAll}(B_i, B_{i+1}) = \begin{cases} 1, & \text{if } \Delta\tau = \infty \text{ and } \forall t_m \in T_{i+1} \cap T_i, \\ & \quad B_i(t_m) = B_{i+1}(t_m), \\ 0, & \text{if } \Delta\tau = \infty \text{ and } \exists t_m \in T_{i+1} \cap T_i, \\ & \quad B_i(t_m) \neq B_{i+1}(t_m). \end{cases} \quad (1)$$

In reality, running the entire set of test cases one by one is not feasible. Thus, teams resort to alternatives for RegAll, such as finding a minimum number of tests (regression test minimization, RTM), selecting a subset (regression test selection, RTS), or running tests in a particular order (regression test prioritization, RTP) [13]. This ensures that regression efforts remain practical under time and resource constraints.

Below, we present the RTM, RTS, and RTP definitions using the previous section's extended definitions. All definitions are written as defined by Yoo and Harman 2012 [13]. The definitions often refer to concepts such as "satisfies a requirement," the set of permutations $PT$, or a prioritization function $f$ that maps to real numbers, which have not been rigorously formalized in the sources nor clarified. In order to remain authentic to these standard formulations, the original terminology and problem statements are preserved without introducing additional formal details or modifications. These definitions are embedded into the notational framework ($B_i$, ($T_{i+1} \cap T_i$), ($S_i \cap S_{i+1}$)) to align with agile workflows and remain consistent with established treatments in the literature.

*Regression Test Minimization (RTM)*: Given a test suite ($T_{i+1} \cap T_i$) and a set of test requirements ($S_i \cap S_{i+1}$) = $\{s_1, \ldots, s_n\}$ that must be satisfied to provide the desired "adequate" testing of the program, let $\{T_{s_1}, \ldots, T_{s_n}\}$ be collections of subsets of ($T_{i+1} \cap T_i$), one subset $T_{s_j}$ for each requirement $s_j$, such that any single test case $t_m \in T_{s_j}$ fulfills $s_j$. The problem is to find a representative set $(T_{i+1} \cap T_i) \subseteq (T_{i+1} \cap T_i)$ that satisfies all of the requirements ($S_i \cap S_{i+1}$); formally, for every $s_j \in (S_i \cap S_{i+1})$, there exists some $t_m \in (T_{i+1} \cap T_i)'$ with $t_m \in T_{s_j}$.

*Regression Test Selection (RTS)*: Given: The original program version $B_i$, the modified version $B_{i+1}$, and a test suite $(T_{i+1} \cap T_i)$. Problem: Find a subset $(T_{i+1} \cap T_i)' \subseteq (T_{i+1} \cap T_i)$ with which to test $B_{i+1}$.

*Regression Test Prioritization (RTP)* Given: A test suite $(T_{i+1} \cap T_i)$, the set of permutations of $(T_{i+1} \cap T_i)$, denoted by $\text{Perm}(T_{i+1} \cap T_i)$, and a function $f_{eval}: \text{Perm}(T_{i+1} \cap T_i) \to R$. Problem: Find a permutation $(T_{i+1} \cap T_i)' \in \text{Perm}(T_{i+1} \cap T_i)$ such that for all $(T_{i+1} \cap T_i)'' \in \text{Perm}(T_{i+1} \cap T_i)$ with $(T_{i+1} \cap T_i)'' \neq (T_{i+1} \cap T_i)'$, $f((T_{i+1} \cap T_i)') \geq f((T_{i+1} \cap T_i)'')$.

In summary, the formal definition presented in this section reduces the long-standing gap between the classical two-build view of regression testing and the continuous, time-bounded reality of agile delivery. Treating RegAll (the ideal case with unlimited time) and the minimization, selection, and prioritization techniques with the same mathematical notations creates a single vocabulary in which every regression-testing decision can be stated, analyzed, and compared. The fact that the well-established RTM, RTS, and RTP problems slot into the framework without alteration shows that the formulation is not an extra layer of notation but a common foundation broad enough to accommodate accepted practice. Later sections demonstrate that this foundation lets practitioners and researchers define any agile regression problems with precision for comparison and analysis

## 5 Validation, Soundness and Completeness

In this section we present the validation of our extended formal definitions using tow state of the art algorithms of regression testing in agile followed by proving the soundness and completeness of our definitions.

### 5.1 Validation

This section validates that the proposed formalism can be used to represent prior works in agile regression testing. We selected two leading agile regression testing approaches: the RETECS method by Speiker et al. [7] and the approach by Bertolino et al. [8]. The mappings between the parameters used in these approaches and their corresponding terms from our extended definition are presented in Tables 1 and 2 respectively. This is followed by an overview of each approach and its representation using our newly defined terms. Note: $t_m.duration = t_m.exectime + t_m.setup$.

*Case 1*

RETECS [7] is an approach for regression testing in agile contexts. The approach works in two steps, namely 1) Time-limited Test Case Prioritization (TTCP) and 2) Adaptive Test Case Selection (ATCS). The mappings in Table 1 map our definitions to RETECS, followed by Algorithm 1, which showcases the complete representation of RETECS with our extended definitions. All inputs and outputs needed to represent RETECS have been listed in Table 1, except for the quality metric and test case execution time. This work uses the quality metric notation $Q$ as Spieker et al. (2017). The quality metric can be anything from coverage to fault detection rate, or anything the team or organization sees fit (this work does not focus on quality metrics and their evaluation). RETECS is a reinforcement learning-based regression test technique that applies model-free, online learning to prioritize and select test cases across continuous integration cycles, adapting dynamically to changing failure patterns. It requires only minimal test metadata, historical verdicts, durations, and last execution times, and does not depend on code coverage, traceability, or static analysis. The method supports discrete tableau-based agents and continuous neural network-based agents, offering scalability via experience replay. Test scheduling is greedy based on learned priorities under a fixed execution time constraint. Adaptation is achieved without pre-training, with convergence to an effective prioritization strategy typically within 60 CI cycles. For more details on RETECS, please refer to the study by Spieker et al. 2017.

**Table 1** RETECS Mappings

| Mappings | RETECS | Proposed-Definition |
|---|---|---|
| Initial set of tests | $TS_i$ | $T_i \cap T_{i+1}$ |
| Evaluation Metric | $Q$ | $Q$ |
| Available Time | $M$ | $\Delta\tau$ |
| TTCP Input | $P(TS_i)$ | $P(T_i \cap T_{i+1})$ |
| TTCP Output | $TS'_i$ | $(T_i \cap T_{i+1})'$ |
| ATCS Input | $TS_i, \ldots, TS_{i-1}$ | $(T_{i+1} \cap T_i), \ldots$ |
| ATCS Output | $TS_i$ | $(T_i \cap T_{i+1})$ |

---
**Algorithm 1** RETECS
---
1: **Initialization** with inputs $T_i \cap T_{i+1}$, $Q$, $\Delta\tau$
      **Time-Limited Test Case Prioritization (TTCP)**
2: **for** each permutation $(T_i \cap T_{i+1})'$ of $(T_i \cap T_{i+1})$ **do**
3:    Calculate $Q((T_i \cap T_{i+1})')$
4:    **if** $\sum_{t_m \in (T_i \cap T_{i+1})'} t_m.duration \leq \Delta\tau$ **then**
5:      **break**
6:    **else**
7:      Continue with another permutation
8:    **end if**
9: **end for**
      **Adaptive Test Case Selection (ATCS)**
10: Selects an order/sequence of test cases from previously executed sequences: $(T_{i+1} \cap T_i)_i, \ldots, (T_{i+1} \cap T_i)_{i-1}$ to maximize $Q$ within $\Delta\tau$
11: **Execution and collect outcomes**
12: **Review and move to next build**
---

*Initialization.* Given the current build $T_{i+1}$ and the immediately preceding build $T_i$, the algorithm starts with the regression-test pool $T_i \cap T_{i+1}$. Together with this pool, it receives a quality (or utility) function $Q$ and an execution budget *Delta tau*. These three artifacts constitute the input on which all subsequent decisions are based.

*Time-Limited Test-Case Prioritization (TTCP).* The algorithm next performs an exhaustive enumeration of the permutations of $T_i \cap T_{i+1}$. For each order, it computes the quality score $Q((T_i \cap T_{i+1})')$ and sums the expected execution times of the constituent test cases. The first permutation whose cumulative duration does not exceed the time budget $\Delta\tau$ is accepted as an ordering; once such a feasible ordering is identified, the search terminates immediately to conserve computational effort.

*Adaptive Test-Case Selection (ATCS).* With a feasible order in place, the method inspects the suites executed in earlier cycles $(T_{i+1} \cap T_i)_i, \ldots, (T_{i+1} \cap T_i)_{i-1}$ and greedily selects the order that maximizes the same quality function while ensuring the total runtime of the augmented schedule still respects $\Delta\tau$. This adaptive step opportunistically uses any remaining time after TTCP to enhance overall fault detection capability.

*Execution and Review* The order is executed, producing verdicts, updated execution times, and coverage data. These outcomes are recorded and subsequently reviewed, providing the empirical evidence required to recalibrate $Q$ and guide RETECS when the next build, $T_{i+2}$, becomes available.

In summary, case 1 confirms the reach of the proposed formalism. Every element RETECS relies on its pool of candidate tests, per-build time budget, learned ranking of tests, and rolling state update translated directly into the existing build-tuple vocabulary; no auxiliary variables, ad-hoc predicates, or special rules were needed. That seamless fit shows that formalism is already broad enough to represent a modern reinforcement-learning regression-testing technique that is exactly as practiced.

## *Case 2*

**Table 2** Bertolino et al. Mappings

| Mappings | Bertolino et al. | Proposed-Definition |
|---|---|---|
| Initialization | *AllCommits* | $B_n$ |
| Evaluation Metric | $Q$ | $Q$ |
| Available Time | $N$ | $\Delta\tau$ |
| Selection Input | DepG of commits | DepG of $(B_i, B_{i+1})$ |
| Selection Output | $(TB_i)$ | $(T_i \cap T_{i+1})$ |
| Prioritization Input | $(TB_i)$, test's history | $(T_{i+1} \cap T_i)$, test's history |
| Prioritization Output | $(TB_i)'$ | $(T_i \cap T_{i+1})'$ |

Bertolino et al. [8] also describe a two-step process (selection, then prioritization), but they do not provide explicit mathematical notation. In Table 2, we introduce the formal symbols to represent each concept from its workflow. Under "Bertolino et al.," `AllCommits` indicates the entire commit history, mapped to $B_n$. Their evaluation metric $Q$ is retained as the one we used in RETECS. The parameter `N`, used to indicate the available time or resource constraint, is expressed as $\Delta\tau$ in our notations.

For selection, Bertolino et al. refer to the `DepG of commits` (the dependency graph that identifies the impacted classes), which we map to DepG($B_i, B_{i+1}$). The output of that selection step, ($TB_i$)′, becomes ($T_i \cap T_{i+1}$)′ in our definitions. Likewise, for prioritization input, ($TB_i$)′ +test's history is represented as ($T_i \cap T_{i+1}$), test's history, while the final prioritized set ($TB_i$)′ is mapped to ($T_i \cap T_{i+1}$)′. Bertolino et al. define prioritization primarily in terms of maximizing early fault detection and minimizing test execution time, using failure history and execution metadata from the last few commits without requiring live instrumentation or runtime tracing. Their dependency graph construction is static and class-level, deliberately avoiding dynamic or coverage-based methods due to scalability constraints in continuous integration. They propose machine learning-based prioritization but leave the choice of ranking models flexible without enforcing a single algorithm or fixed policy. By explicitly defining these terms, we ensure the Bertolino et al. approach can be uniformly interpreted and integrated with other mathematically defined methods. For more details on the original steps, see Bertolino et al. [8].

---

**Algorithm 2** Bertolino et al.

1: **Initialization** with inputs $B_n$, $Q$, $\Delta\tau$
   **Test Selection**
2: Build initial class-level dependency graph DepG from $B_n$
3: **for** each CI cycle with commits ($B_i, B_{i+1}$) **do**
4:    Update DepG considering changed classes between ($B_i, B_{i+1}$)
5:    ($T_i \cap T_{i+1}$) ← DepG($B_i, B_{i+1}$)
   **Test Prioritization**
6:    ($T_i \cap T_{i+1}$)′ ← Order[($T_i \cap T_{i+1}$), history data] to maximize $Q$ AND $t_m.duration \leq \Delta\tau$
7:    **Execute and collect outcomes**
8: **end for**
9: **Review and move to next build cycle**

---

*Initialization.* The procedure receives the current build $B_n$, a quality/utility metric $Q_i$, and a global time budget $\Delta\tau$. From $B_n$, it constructs an initial class-level dependency graph *DepG* that links each test case to the production classes it touches.

*Test Selection.* For every subsequent continuous integration cycle that transforms build $B_i$ into $B_{i+1}$, the approach incrementally updates *DepG* to reflect the classes changed by the latest commits. Querying the revised graph isolates the tests affected by those modifications, yielding the candidate set $T_i \cap T_{i+1}$.

*Test Prioritization.* Leveraging historical execution data, an *Order* routine/function ranks the candidate tests. It prioritizes tests based on maximizing $Q$ and whose cumulative estimated runtime does not exceed the budget $\Delta\tau$, producing an order ($T_i \cap T_{i+1}$)′.

*Execution and Review.* The prioritized suite is executed on $B_{i+1}$; verdicts, actual timings, and coverage are recorded and fed into the history. After a review, $B_{i+1}$ the cycle restarts with the refreshed artifacts *{DepG, Q,* history*}*.

In summary, case 2 shows that the dependency-graph approach proposed by Bertolino et al. also drops straight into our formalism. It is an evolving graph, the time box, the ranked list of affected tests, and the incremental update after each commit. All maps are mapped onto elements already present in the build-tuple language, so no additional symbols or special rules are required. That seamless fit confirms the framework's breadth: it accommodates a graph-driven selection strategy as naturally as it handled the learning-based example in case 1, strengthening the claim that the formalization is a common foundation for diverse agile regression-testing techniques

## 5.2 Soundness and Completeness

With the above-mentioned representations of both Spieker et al.'s CI-scheduling method and Bertolino et al.'s dependency-graph algorithm, it was shown to instantiate our model via build sequences $B_1, B_2, \ldots B_n$. where each choice and ordering of tests in the overlap $T_i \cap T_{i+1}$ exactly mirrors that strategy's test-selection and prioritization of its CI environment under a real budget $\Delta\tau$ and monotonic scope function $f_{\Delta\tau}$, and its quality metric $Q$. These two mappings cover the primary paradigms of agile regression testing—time-boxed setting demonstrating the expressiveness of our Build-tuple formalism.

Showing that two algorithm fit the model establishes flexibility, but to establish completeness and soundness requires proving that every agile regression-testing strategy applied on one branch of the development can be embedded similarly. By branch, we mean a parallel paths of development where changes can be made independently of the main sequence of builds that goes to production. Without this generality claim, the framework might be considered valid only for the cases studied rather than as a foundational theory.

***Scope:*** Throughout this section we consider a *single, linear* stream of builds indexed by $i = 1, 2, 3,$ Each build has exactly one wall-clock interval $\Delta\tau_i$. No claim is made about multi-branches. The quality indicator $Q$ is intentionally unspecified; it may be any of the agile regression testing metrics such as APFD, test coverage, cost etc or a blend of blend of such metrics [9].

**Definition 1:** *Agile Regression-Testing Algorithm (ARTA) - For every build number $i \in \{1, 2, 3,$  $\}$ an ARTA:*

1. receives a program version $P_i$, an active requirement set $S_i$ and a candidate test set $T_i$; each of these fall in the area 'Set of Interest' [Figure 2]
2. selects a subset of $T_i$ or ordered for prioritization and/or trimmed for minimization based on the $Q$
3. finishes executing that subset within a finite wall-clock interval $\Delta\tau_i$ that ends before build $i + 1$ begins.

*Note:* No restriction is placed on the algorithm's mechanics.

**Proposition 1** *Let R be any* ARTA. *There exists a* constructive *mapping*
$$\Psi_R : i \longrightarrow (P_i, S_i, T_i, \Delta\tau_i, Q) \quad (i = 1, 2, 3, \ldots\ldots)$$

*such that, for every build i, the 5-tuple $\Psi_R(i)$ records* exactly *the test subset with any ordering or minimization executed by R at build i, together with the time budget $\Delta\tau_i$ and the quality value Q employed by R.*

**Table 3** Constructive mapping

| Tuple field in $\Psi_R(i)$ | Corresponding artifact at build $i$ |
|---|---|
| $P_i$ | Program version |
| $S_i$ | Requirement / specification set |
| $T_i$ | Complete test repository |
| $\Delta\tau_i$ | Finite wall-clock interval declared by $R$ |
| $Q_i$ | Quality value produced or optimized by $R$ |

The completeness claim is realized by extracting, for every build $i$, the five elements listed in Table 3. Collecting these tuples for $i = 1, 2, 3, \ldots$ yields the mapping $\Psi_R$ required by Proposition 1. Let us assume there exists an ARTA $R^*$ such that no mapping $\Psi_{R^*}$ can represent its behavior on all builds $1, \ldots, k$. Then

$$\exists j \in \{1, \ldots, k\} \text{ s.t. } \Psi_{R^*}(j) \not= \langle P_j, S_j, T_j, \Delta\tau_j, Q \rangle.$$

- By Definition 1, for build $j$:
  - The artifacts $P_j, S_j, T_j, \Delta\tau_j$ exist.
  - $R^*$ selects and/or orders tests satisfying the time budget $\Delta\tau_j$ and computes the score $Q$.
- Explicitly construct for each $i = 1, \ldots, k$:

$$\Psi_{R^*}(i) := \langle P_i, S_i, T_i, \Delta\tau_i, Q \rangle.$$

  By construction, $\Psi_{R^*}(j)$ exactly matches $\langle P_j, S_j, T_j, \Delta\tau_j, Q \rangle$.
- This contradicts the assumption made previously.
- Therefore, no such $R^*$ exists, and every ARTA $R$ within the scope admits a tuple mapping $\Psi_R$ that completely represents it over builds 1 through $k$.

Another significant corollary is that the build-tuple formalism also subsumes classical two-version regression testing as a special case. If one restricts attention to a single pair of builds $\langle P_i, S_i, T_i \rangle \longrightarrow \langle P_{i+1}, S_{i+1}, T_{i+1} \rangle$ and allows $\Delta\tau \to \infty$, the model collapses to the traditional paradigm of executing $T_i \cap T_{i+1}$ in its entirety—a process long familiar from the work of Rothermel and Harrold [6]. In much the same way that oracle-based comparisons treat the earlier version as a static baseline [17], our formalism generalizes and extends this idea into a continuous chain. Previously validated behaviors in $B_i$ become the reference for $B_{i+1}$, and the same comparison logic repeats across an unbounded sequence of builds. Thus, the fundamental principle of "compare-and-verify" remains intact yet is situated within an iterative, resource-aware structure that mirrors real-world agile processes.

In conclusion, we have constructed an explicit, algorithmic mapping $\Psi_R$ that applies to *every* regression-testing strategy operating on a single build branch. The step-by-step construction, anchored in Table 3, shows precisely how to lift a strategy's artifacts—program version, requirement set, tests, time budget, and quality metric into a five-field tuple for each build, while the accompanying contradiction argument confirms that no strategy satisfying our scope can evade that representation. Together, the soundness (constructive recipe), completeness proof, and collapse to one pair of builds demonstrate that the build-tuple framework captures the space of single-branch agile and non-agile/traditional regression testing algorithms: regardless of how an algorithm defines versions, requirements, tests, wall-clock limits or quality metrics, it can be expressed without loss in this formalism.

# 6 Discussion

In this work, we have developed a unified, mathematically rigorous framework for regression testing tailored to the realities of agile and continuous-integration environments. Moving beyond the classical view of regression testing as a one-off comparison between two static program versions, our model treats software evolution as a continuous sequence of build tuples, each of which bundles together the program snapshot, its current set of user-story specifications, and its complete regression-test suite. By introducing an explicit regression-test window, a real-valued time interval between successive builds, and by associating it with a monotonic scope function that quantifies how many of the overlapping tests can execute under given resource constraints, we provide a precise mechanism for reasoning about test-execution scope across commit-level, nightly, sprint, and release cycles. It captures modern development teams' operational pressures, where time budgets fluctuate and requirements evolve on every build. In doing so, we fulfill the primary research aim set out in the introduction i.e. to provide a formal framework for regression testing in agile environments thereby translating an industry problem into a mathematically tractable one.

We validated this formalism's expressiveness and practical adherence through detailed mappings of two state-of-the-art agile regression-testing techniques: the RETECS method and Bertolino et al. In both cases, the underlying algorithms reduce to the single abstract operation of selecting and ordering tests in the shared subset of successive suites, constrained by the regression-test window and guided by a quality metric. This correspondence demonstrates that our framework is not merely a theoretical construct but a reliable framework for capturing real-world CI workflows. Building on the concrete embeddings of RETECS and Bertolino et al., Section 5 establishes by proof that every agile regression-testing algorithm within our defined scope can be written directly in the build-tuple notation. Each algorithm's essentials are its notion of versions, its set of tests, requirements, and its time budget, which can be mapped to fields in the tuple. Because this translation loses no information, running the tuple-based rendition reproduces the original algorithm's outcomes exactly. Practically, that guarantee allows teams to replay, compare, or swap algorithms within a CI pipeline without altering visible behavior while providing researchers with a single, uniform language for deeper analysis.

The implications of this unified framework extend well beyond theoretical elegance. For practitioners, structuring CI/CD pipelines around build tuples provides scope for collecting data on the regression testing process to make clear, data-driven decisions for test planning and execution. With these elements formalized, teams can automate the computation of the optimal subset of tests to run, balancing defect-detection coverage against release deadlines. They can adjust strategy dynamically as project priorities shift. Tool vendors gain a standardized abstraction layer: new test-ranking modules or scheduling policies can be introduced simply by defining corresponding functions within the existing formal semantics, without rewriting pipeline orchestration code. At the organizational level, the framework offers a common language for auditing and comparing regression-testing regimes across teams, facilitating knowledge transfer, identifying bottlenecks, and driving continuous improvement.

In academic research, the build-tuple model establishes a shared notation and set of primitives for describing, analyzing, and comparing regression-testing techniques. Researchers proposing novel selection and prioritization algorithms need only specify how their method fits the model's parameters. Meta-analytic studies become more tractable, as results from heterogeneous experiments can be aggregated when every approach reduces to the same formal operations on build tuples. This framework also opens the door to deeper theoretical studies. For example, researchers could explore how the function determining how many tests fit into a time window behaves when the deadline is uncertain, when tests have different execution times, or when the aim is to minimize the overall cost. They could also analyze whether different ways of ranking tests (by failure history, coverage, or other metrics) reliably lead to the best outcomes as more builds are processed.

The framework also unlocks novel lines of cross-disciplinary inquiry, including business-value annotations in the specification, and positions regression testing as an inherently value-driven activity, bridging the gap between technical validation and strategic decision-making; the metric that must be used to measure regression testing in agile as recently pointed by Das and Gary [9]. Economists and management scientists could model regression testing as a cost-benefit problem, navigating trade-offs between defect-detection coverage and time-to-release. Operations researchers might adapt scheduling algorithms from real-time systems to optimize test ordering under variable budgets. Machine-learning experts could leverage historical build tuples as feature sets for predictive models that forecast test outcomes or dynamically adjust time allocations. In each of these areas, the build-tuple abstraction provides an explicit mapping between domain-specific concepts and the formal parameters of the model.

Building on this foundation, our framework lays the groundwork for several important extensions. Incorporating branching and merging workflows would require generalizing build sequences to partially ordered sets and capturing parallel development streams and complex integration topologies. Modeling environment variabilities such as hardware and human-in-the-loop dependencies would also be necessary. Empirical validation at scale, involving the analysis of real build histories and test logs from large-scale industrial projects, will refine assumptions, making this model much richer.

In summary, this work transforms regression testing from a collection of empirical heuristics into a coherent theory by delivering a complete, rigorous foundation for regression testing in agile and CI environments. The build-tuple model is the definitive reference for researchers and practitioners, providing the semantics needed to describe, compare, optimize, and extend regression-testing strategies in the face of accelerating development cycles. We therefore invite practitioners and researchers alike to operationalize, benchmark, and evolve the build-tuple abstraction within their own CI pipelines. Doing so will transform the formalism presented here into a living standard that keeps pace with the cadence of modern software delivery.

## 7 Limitation and Future Works

Our framework models regression testing as a linear sequence of builds $B_1, B_2, \ldots, B_n$, each combining program code, specifications, and test suites. However, In large-scale projects where testing is a challenge irrespective of the line of business, the project handles [18], [19], development proceeds across multiple parallel branches feature branches, defect fix streams, and long-lived release lines that merge irregularly. Capturing these workflows requires extending the model from a simple chain to a directed acyclic graph of builds, with formally defined merge points and test-set resolution rules when branches converge. Without this generalization, our framework cannot represent scenarios where tests must validate cross-branch interactions before or after merges.

The current definitions treat the test environment as implicit: each build assumes a fixed execution context. In practice, container images, infrastructure-as-code scripts, and external service configurations evolve alongside code, and these changes can introduce regressions or spurious failures. Future work should extend each build tuple to include an explicit environment descriptor capturing container tags, configuration versions, and service endpoints and refine the regression-test window to account for environment-driven setup overhead and variability in test outcomes under different configurations.

We focus on functional regression tests, leaving non-functional concerns such as performance benchmarks, security scans, and load tests outside the formalism. However, CI pipelines routinely interleave these test classes, allocating separate or shared time budgets and applying different pass/fail criteria. Incorporating multiple test dimensions would involve partitioning each test suite $T_i$ into categorized subsets (functional, performance, security, etc.) and defining composite scope functions that allocate the overall time window $\Delta\tau$ among them according to risk profiles or service-level agreements. This extension is necessary for teams that balance functional correctness with performance targets and compliance requirements within the same pipeline.

Our binary view of test outcomes does not account for defect severity or repair complexity. In practice, teams prioritize tests by their likelihood of failure and the downstream cost and risk associated with fixing high-impact defects. Future work should introduce a defect-impact function such as $D(t_m)$ for each test case $t_m$ and model repair time, enabling joint optimization of test selection under dual detection and remediation effort constraints. This addition would allow the framework to recommend test schedules that minimize project risk and resource expenditure.

Moreover, modern software systems increasingly embed continuously trained machine-learning components whose behavior depends not only on the source code but also on mutable data and retraining schedules [20]. MLOps practice extends the CI/CD loop with Continuous Training and adds dedicated data manipulation, model creation, and model deployment pipelines. These pipelines introduce regression-testing challenges that our current build-tuple model does not yet capture: data-driven drift, non-deterministic model weights, and validation workflows that must re-verify both code and model artifacts after each training cycle. Extending the framework will, therefore, require things such as (i) versioning of datasets and trained models within each build tuple and (ii) a time window that distinguishes retraining from inference validation.

Finally, our framework rests on a theoretical construct whose shapes and parameters have not been calibrated against industrial data. Rigorous empirical validation is needed: analyzing real build histories, test-execution logs, and defect-fix records from diverse organizations will reveal whether the assumptions of linear build sequences, fixed test durations, and utility-driven ranking hold in practice. Longitudinal case studies with industry partners can evaluate how well the model guides actual test-planning decisions. Such collaborations will also uncover additional factors, such as dynamic iteration lengths and evolving organizational processes, that must be folded into the formalism to ensure its relevance and usability.

Addressing these limitations, parallel branch modeling, explicit environment descriptors, multi-dimensional test categorization, defect-impact modeling, and empirical calibration will transform our initial theory into a comprehensive framework. By formally incorporating these elements, future work can deliver a context-aware model that supports the full complexity of agile, CI-driven development and provides actionable guidance for researchers and practitioners.

## 8 Additional Information

- **Corresponding author:** Suddhasvatta Das, email: sdas76@asu.edu
- **Author Contribution(s):**
  - Suddhasvatta Das (first author) - is responsible for the entire work from ideation to writing
  - Dr Kevin Gary (second author) - is responsible for reviewing the entire work

- **Data Availability:** This work is purely theoretical (mathematical) in nature. Thus, there is no such 'data' that is required to understand and/or replicate this work.
- **Funding:** This research did not receive any specific grant from funding agencies.
- **Competing Interest:** The authors declare that they have no conflict of interest.

**Acknowledgments.**

- This work acknowledges the assistance of a Large Language Model (LLM) used explicitly for editing/drafting purposes. The author has reviewed all content to ensure accuracy, integrity, and correctness.